\documentclass[revtex4]{emulateapj}

\usepackage{amssymb}
\usepackage{amsmath}
\usepackage[]{graphicx}
\usepackage{enumerate}
\usepackage{subeqnarray}
\usepackage{cases}
\usepackage{mathrsfs,amssymb}
\usepackage[usenames]{color}

\tightenlines

\begin{document}

\title{Magnetic mirroring and focusing of cosmic rays}

\author{Kedron Silsbee$^1$, Alexei V. Ivlev$^1$, Marco Padovani$^2$, Paola Caselli$^1$}
\email[e-mail:~]{ksilsbee@mpe.mpg.de} \email[e-mail:~]{ivlev@mpe.mpg.de} \affiliation{$^1$Max-Planck-Institut f\"ur
Extraterrestrische Physik, 85748 Garching, Germany } \affiliation{$^2$INAF--Osservatorio Astrofisico di Arcetri, Largo E.
Fermi 5, 50125 Firenze, Italy}

\begin{abstract}
We study the combined impact of magnetic mirroring and focusing on the ionization by cosmic rays (CRs) in dense molecular
clouds and circumstellar disks. We show that for effective column densities of up to $\sim10^{25}$~cm$^{-2}$ (where
ionization is the main mechanism of energy losses by CRs) the two effects practically cancel each other out, provided the
magnetic field strength has a single peak along field lines. In this case the ionization rate at a given location is
controlled solely by attenuation of interstellar CRs due to energy losses. The situation is very different in the presence
of magnetic pockets -- local minima of the field strength, where the CR density and thus ionization can be reduced
drastically. We obtain simple analytical expressions allowing accurate calculation of the ionization rate in these regions.
\end{abstract}


\maketitle

\section{Introduction}
\label{sec:intro}

The ionization degree of molecular clouds is a critical factor in the dynamics of star formation. A small fraction of
ionized species controls the coupling of the Galactic magnetic field to the predominantly neutral gas of the cloud,
influencing its stability against gravitational collapse \citep[][]{Mestel1956}, the efficiency of the fragmentation process
\citep[][]{Price2008}, and the formation of circumstellar disks around young stars \citep[][]{Allen2003}. The main source of
ionization in dark regions of molecular clouds and pre-stellar cores is cosmic rays (CRs), which initiate a chain of
chemical reactions starting from the collisional ionization of the most abundant species, molecular hydrogen
\citep[][]{Yamamoto2017book}.

CRs, responsible for the ionization in dense cores and circumstellar disks, propagate along the local magnetic field. The
magnetic configuration in such objects can be very complicated \citep[][]{Joos2012,Li2013,Padovani2013}, and the field
strength can be much larger than the interstellar value \citep[][]{Crutcher2012}. The field strength increases along the
field lines converging into denser central regions, which leads to efficient mirroring of the penetrating CRs -- their pitch
angles increase in response to the growing field until reaching $90^\circ$, and thus more and more particles are reflected
back. On the other hand, the convergence of field lines results in the CR focusing. These two competing effects play
important roles in various processes occurring in molecular clouds
\citep[][]{Cesarsky1978,Ko1992,Chandran2000,Desch2004,Padoan2005}.

Recently, there have been studies \citep[][]{Padovani2011,Padovani2013} investigating the combined effect of magnetic
mirroring and focusing on the CR ionization in the dense molecular cores. A comprehensive analysis of the CR propagation in
static and collapsing magnetized clouds has been carried out, by varying the relative strength of the toroidal/poloidal
field components and the mass-to-flux ratio. The authors concluded that mirroring always dominates over focusing, implying a
reduction of the CR ionization rate by a factor of about 2--3 with respect to the case where these magnetic effects are
neglected. It was also shown that, for large values of the flux-to-mass ratio, these effects reduce the ionization in
collapsing clouds by more than an order of magnitude, which can have important consequences for the dynamical evolution and
the formation of disks \citep[e.g.,][]{Zhao2018}.

The major aim of the present work is to perform a general analysis of the effects of CR mirroring and focusing in
dense cores and circumstellar disks. We identify universal mechanisms that govern the CR ionization in such environments, assuming that there is no stochastic change in the pitch angle. This assumption is well justified, since the field
structure normally remains stationary at a timescale of CR crossing (since the physical velocity of CRs is typically a few
orders of magnitude larger than the velocity of MHD waves).\footnote{Fast processes associated with possible magnetic
reconnection \citep[see, e.g.,][]{Lazarian2014}, induced in these dense regions by rotationally-driven MHD turbulence,
require separate consideration which is beyond the scope of this paper.}
We also make use of the fact that the small-scale resonant MHD turbulence, which could lead to efficient CR
scattering, is completely damped under such conditions due to frequent ion-neutral collisions \citep[e.g.,][]{Ivlev2018}.

In Section~\ref{noLoss} we study the net effect of the mirroring and focusing on the CR density when energy losses are
negligible; we identify two distinct cases, of single- and multiple-peaked magnetic field strength, and demonstrate that in
the former case the mirroring and focusing cancel out exactly, while in the latter case a drastic reduction of the CR
density is possible. In Section~\ref{ionLoss} we determine the exact upper and lower bounds for the ionization rate in the
two cases; we show that ionization is practically unaffected for a single-peaked field and is reduced for a multiple-peaked
field, and provide analytical expressions for the ionization rate. Finally, in Section~\ref{discussion} we summarize our
main findings and briefly discuss implications for the relevant astrophysical problems.

\section{Mirroring and focusing without losses}
\label{noLoss}

\begin{figure}[htp]
\centering
\includegraphics[width=1.03\columnwidth]{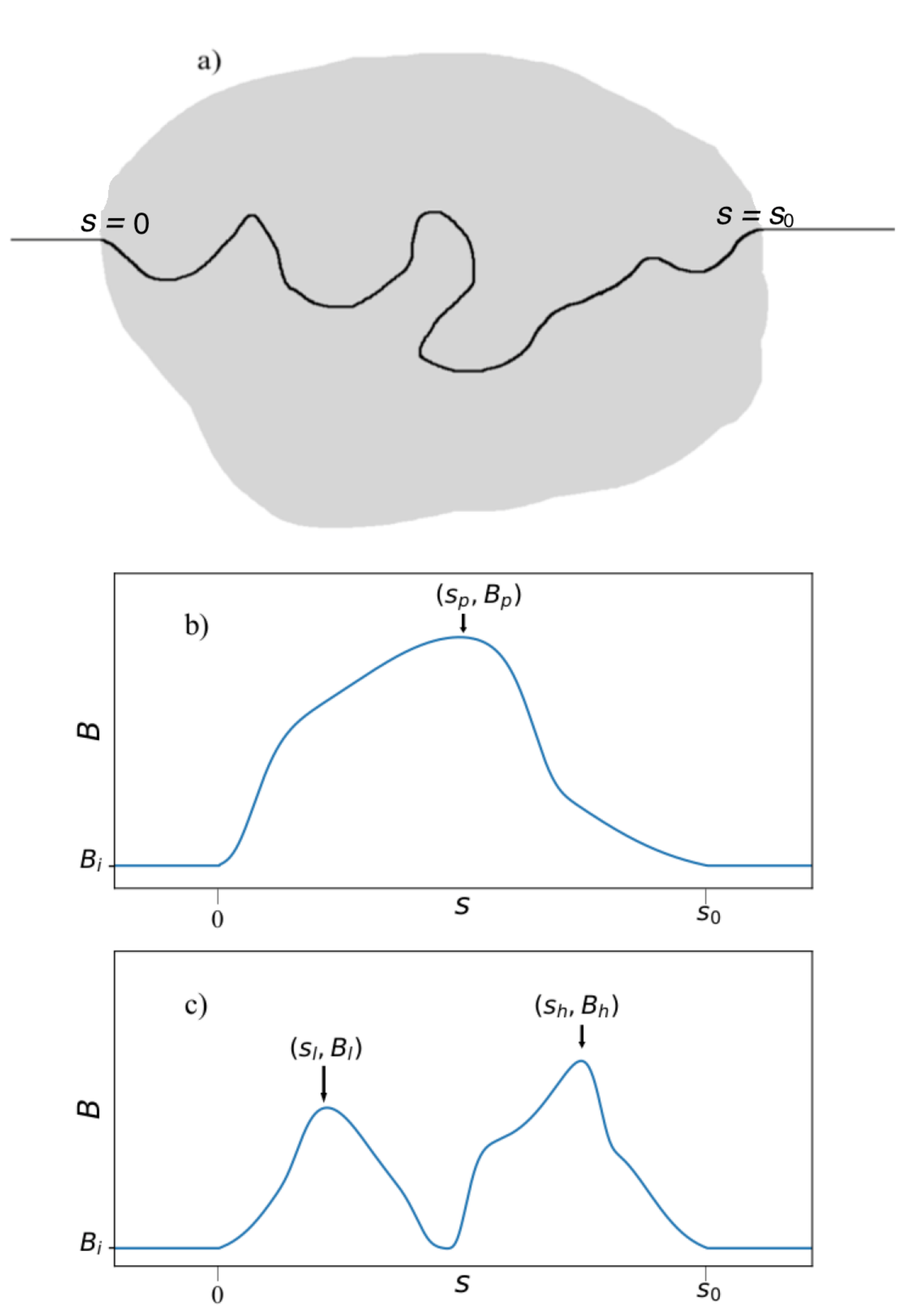}
\caption{Panel (a) shows a sketch of a magnetic field line entering a cloud (grey region). Panels (b) and (c)
depict sketches of the magnetic field strength $B$ as a function of distance $s$ along a field line. Panel (b) shows the
single-peaked case discussed in Section~\ref{singleMaximum}, panel (c) shows the double-peaked case,
Section~\ref{multipleMaxima}.} \label{introFig}\vspace{.1cm}
\end{figure}

Consider the magnetic field threading a molecular cloud or circumstellar disk (below we use the term ``cloud''), as sketched
in Figure~\ref{introFig}(a). Let $s$ be the distance coordinate along a field line, and assume that outside the cloud there
is a constant (interstellar) field strength $B_i$. The cloud is surrounded by the interstellar medium with isotropically
distributed CRs that can travel in both directions without energy losses, along any field line which penetrates the cloud.

In the absence of scattering processes, the pitch angle $\alpha$ (between the velocity of a CR particle and the field)
satisfies the relation
\begin{equation}
\frac{\sin^2\alpha}{B(s)}=\frac{\sin^2\alpha_i}{B_i}\;,
\label{mirroringEquation}
\end{equation}
which follows from the adiabatic invariance of the magnetic moment of a particle \citep[e.g.,][]{Chen1984book}. In this
case, the steady-state kinetics of CRs with given momentum $p$ is characterized by the distribution function
$f(\mu,p,s)$,\footnote{For brevity, below we only show arguments of $f$ that are essential for understanding. We use the
normalization such that total number density of CRs is $\int dp\int f\:d\mu$.} obeying the equation
\citep[][]{Morfill1976,Cesarsky1978}
\begin{equation}\label{FP1}
\mu\frac{\partial f}{\partial s}-(1-\mu^2)\frac{d\ln \sqrt{B}}{ds}\frac{\partial f}{\partial\mu}=0,
\end{equation}
where $\mu\equiv\cos\alpha$. A general solution of this equation is any function of $(1-\mu^2)/B(s)$, showing that the CR
distribution is conserved for a given magnetic moment. This conclusion, which naturally follows from the Liouville theorem,
implies
\begin{equation}\label{conservation1}
f(\mu,s)=f_i(\mu_i),
\end{equation}
where $f_i(\mu_i)$ is the distribution function of interstellar CRs. The local value of $\mu$ is determined from
Equation~(\ref{mirroringEquation}),
\begin{equation}\label{cos_alpha}
\mu(\mu_i,s)=\pm\sqrt{1-\tilde B(s)(1-\mu_i^2)}\;,
\end{equation}
where $\tilde B(s) = B(s)/B_i$ is the magnetic ``focusing factor'' (increasing $B$ implies the proportional focusing of
field lines). A monotonic increase of the magnetic field strength leads to the particle mirroring: interstellar CRs can
reach the position $s$ only if $\sin\alpha_i\leq1/\sqrt{\tilde B(s)}$, i.e., if $|\mu_i|\geq\sqrt{1-1/\tilde B(s)}$.

Equation~(\ref{conservation1}) has a simple physical meaning. Using the relation
\begin{equation}\label{dmu}
\frac{\partial\mu}{\partial\mu_i}=\tilde B(s)\frac{\mu_i}{\mu}\;,
\end{equation}
we obtain $f(\mu,s)\mu \:\delta\mu=\tilde B(s)f_i(\mu_i)\mu_i \:\delta\mu_i$. Written in this form, the equation expresses a
conservation of the {\it differential flux} along a field line, with taking into account that the pitch angle varies with
$s$ in accordance with Equation~(\ref{mirroringEquation}) and that the local particle density is proportional to the
focusing factor $\tilde B(s)$.

\subsection{Single Maximum of the Field Strength}
\label{singleMaximum}

In this section we assume that $B(s)$ has only one peak on the interval $0\leq s\leq s_0$, located at $s_p$, as shown in
Figure~\ref{introFig}(b). Consider CRs at $0<s<s_p$ moving in the $+s$ direction (referred to as {\it forward-moving
particles}), and denote their local distribution by $f_+$. Since $B$ (and thus $\alpha$) continuously increase for such
particles, $f_+$ is nonzero for all values of $\mu$ between 0 and 1.

Let us calculate the local differential density (per unit momentum) of forward-moving CRs, $n_+(p,s)=\int_0^1f_+\:d\mu$.
Using Equation~(\ref{conservation1}), we substitute $f_+=f_i$ and keep in mind that the distribution of interstellar CRs is
isotropic, $f_i=\frac12n_i$, where $n_i(p)$ is their differential density. Then the integration over $\mu$ yields
$n_+=\frac12n_i$. Thus, the density of forward-moving particles remains constant and equal to the density of CRs entering
the cloud from the left.

Local CRs also include particles which are moving in the $-s$ direction ({\it backward-moving particles}). Their
distribution $f_-$ is a sum of two components: particles with $-\mu_p<\mu<0$, which were mirrored before reaching the peak
at $s=s_p$, and particles with $-1\leq\mu<-\mu_p$, which passed through the peak from the other side.\footnote{The value of
$\mu_p=\sqrt{1-B/B_p}$ is derived from Equation~(\ref{cos_alpha}) by substituting $1-\mu_i^2=1/\tilde B_p$; the latter
follows from the mirroring condition ($\mu=0$) at the peak.} Taking into account that $f_-(\mu,s)=f_+(|\mu|,s)$ for the
mirrored particles and using Equation~(\ref{conservation1}) for both components, we obtain $f_-=f_i$ and hence $n_-=
\frac12n_i$.

We conclude that the effects of magnetic mirroring and focusing cancel exactly when energy losses by CRs are neglected
(i.e., when Equation~(\ref{FP1}) holds). The distribution of particles does not depend on the location in a cloud and
coincides with the distribution of interstellar CRs, $f_i$ \citep[][]{Cesarsky1978}. For non-relativistic CRs, the latter is
isotropic to a very high degree, and therefore the local density is equal to the density of CRs outside the cloud, $n_i$. We
note, however, that if $f_i(\mu_i)$ exhibits anisotropy \citep[][]{Parker1963book}, the density is not conserved -- it
generally increases (decreases) with $s$ if $f_i$ has a maximum (minimum) at $|\mu_i|\approx1$: For example, a beam
concentrated near $|\mu_i|\approx1$ is practically not mirrored, so the focusing leads to the density increase; on the
contrary, particles with a completely depleted distribution near $|\mu_i|\approx1$ are totally mirrored when they reach a
location where the field is sufficiently strong.

\subsection{Multiple Maxima}
\label{multipleMaxima}

If the magnetic field strength has more than one maximum along a field line, then the focusing factor is still $\tilde
B(s)$, but the calculation of the mirroring effect is different. Therefore, the particle density is no longer constant along
that field line, as shown below.

Suppose first that $B(s)$ has a ``lower'' peak $B_l$ at $s=s_l$ and a ``higher'' peak $B_h$ at $s = s_h$, as in
Figure~\ref{introFig}(c). If either $B(s) \geq B_l$, or $s$ is not between the two peaks, the ``lower'' peak has no effect
on the local particle density, and the results from Section~\ref{singleMaximum} are applicable. A more interesting situation
occurs in a {\it magnetic ``pocket''}, for $s_l < s < s_h$ and $B(s) < B_l$. In this case there are three groups of
particles contributing to the local density: particles moving in the $-s$ direction that came from $s > s_0$ and passed
through the maximum in $B$ at $s_h$; particles moving in the $+s$ direction that came from $s<0$ and passed through the
maximum in $B$ at $s_l$; and finally those which came from $s <0$, passed through $s_l$, but were reflected before reaching
$s_h$ and are now moving in the $-s$ direction.

The contribution of the first (backward-moving) group is calculated by integrating $f_i$ over $-1\leq\mu<-\mu_h$, with
$\mu_h=\sqrt{1-B/B_h}$ (see Section~\ref{singleMaximum}), which yields the density $\frac12n_i(1-\mu_h)$. This result does
not depend on the direction of propagation, and therefore the contribution of the second (forward-moving) group can be
obtained from the same expression by replacing $\mu_h$ with $\mu_l$. The density of the third group of particles is not
affected by their reflection, and thus is calculated by integrating over $\mu_l<\mu<\mu_h$, which yields
$\frac12n_i(\mu_h-\mu_l)$. Adding up the three contributions, we find the local density,
\begin{equation}\label{multiPeak}
\frac{n(s)}{n_i}=1-\sqrt{1 - \frac{B(s)}{B_l}}\;.
\end{equation}
We see that $n /n_i$ does not depend on $B_h$ and is very sensitive to $B_l$.  For example, a 1\% reduction in $B$ from the
``lower'' peak value, i.e., $B/B_l = 0.99$, leads to a 10\% reduction of the local density.

\begin{figure}[htp]
\centering
\includegraphics[width=\columnwidth]{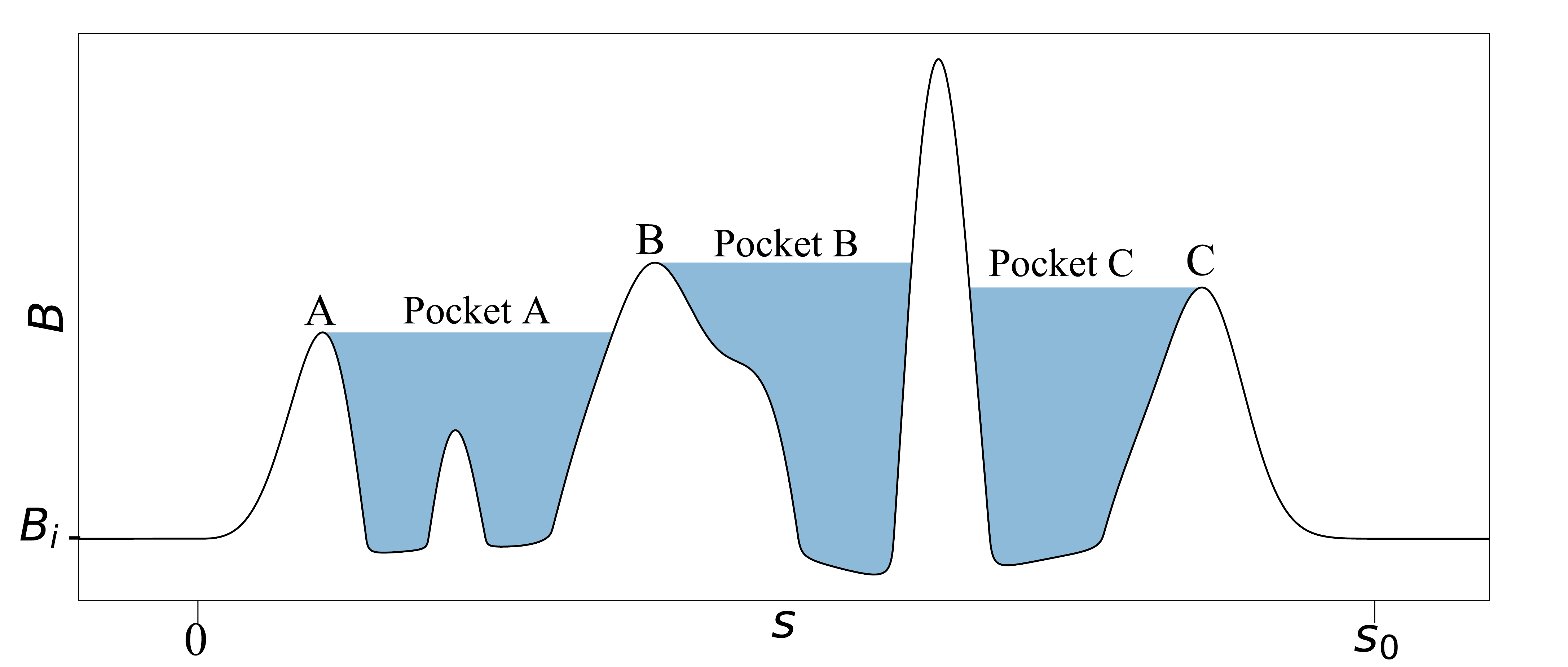}
\caption{A situation where the magnetic field has multiple maxima along the field line. This results in
multiple magnetic pockets, indicated by the shading.} \label{sketch}\vspace{.1cm}
\end{figure}

This result can be straightforwardly generalized for the case of several peaks. Figure~\ref{sketch} illustrates such a
situation, with magnetic pockets indicated by shading. From the same line of arguments as above it follows that the local
density in {\it each} pocket is described by Equation~\eqref{multiPeak}, where $B_l$ should be replaced by the respective
value of the peak field ($B_{\rm A}$, $B_{\rm B}$, and $B_{\rm C}$ for pocket A, B, and C). We notice that
Equation~\eqref{multiPeak} remains applicable even if $B(s)$ has small peaks inside a pocket (as, e.g., for pocket A).

\section{Effect of the mirroring and focusing on the ionization rate}
\label{ionLoss}

In Section~\ref{noLoss} we neglected energy losses by CRs and showed that in this case the magnetic mirroring and focusing
exactly cancel each other, if we assume the field strength has a single peak along field lines. In particular, this implies
that the ionization in a cloud is not influenced by the magnetic effects as long as energy losses do not substantially
attenuate CRs (specifically, do not modify the part of their energy spectrum providing the main contribution to the
ionization, see discussion in Section~\ref{FMP}). Then, irrespective of the strength and configuration of the magnetic field
in the cloud, the ionization rate can be calculated using the density of interstellar CRs (i.e., as if the field strength
inside the cloud remains constant and equal to $B_i$).

When the CR attenuation caused by energy losses cannot be neglected, two factors complicate the simple argument presented
above. On the one hand, the density of forward-moving particles for $dB/ds>0$ becomes {\it higher} compared to that in the
constant-field case: the pitch angle of such particles continuously increases, which means that (for a given value of local
pitch angle) they have shorter helical trajectories and therefore suffer {\it less attenuation}. On the other hand, the
density of backward-moving mirrored particles is reduced at a given point, as they have traveled through more column than
those that are still forward-moving. Furthermore, the density of backward-moving particles passed through the cloud from the
other side is more attenuated than that of forward-moving particles (unless we consider the central region of the cloud,
where mirroring is less important and particles come from both sides with a similar attenuation).

To quantify the effect of the mirroring and focusing on the CR ionization, in Section~\ref{FMP} we first put bounds on the
relative increase in the rate of ionization by forward-moving particles, assuming the field strength has a single peak.
Then, in Section~\ref{mirroringSection} we estimate the ratio of the ionization due to the mirrored particles to the
ionization due to the forward-moving particles. Finally, in Section~\ref{pockets} we calculate a reduction of the ionization
rate in magnetic pockets.

\subsection{Ionization by the Forward-Moving Particles}
\label{FMP}

In the presence of continuous energy losses by CRs, the distribution function is described by the following steady-state
kinetic equation \citep[][]{Skilling1976,Cesarsky1978}:
\begin{equation}
v\mu \frac{\partial f}{\partial s}-v(1-\mu^2)\frac{d\ln \sqrt{B}}{ds}\frac{\partial f}{\partial\mu}+
\frac{\partial}{\partial p}(\dot p_{\rm loss}f) = 0,
\label{transportEquation0}
\end{equation}
which is obtained by adding a loss term to Equation~(\ref{FP1}) (multiplied by the particle velocity $v$). The loss term is
characterized by the momentum decrease of a particle per unit time due to inelastic processes, $\dot p_{\rm loss}<0$.

For further analysis it is convenient to use the particle kinetic energy $E$ instead of the momentum $p$. Taking into
account that $dE=v\:dp$, we obtain that the respective distributions are related via $f(\mu,p,s)=v(E)\hat f(\mu,E,s)\equiv
j(\mu,E,s)$, where $j$ is the so-called energy spectrum of CRs (and $\int\hat f\:d\mu$ is the differential density per unit
energy). One can write the losses due to interaction with gas particles in the form $\dot p_{\rm loss} = -n_g(s)L(E)$, where
$n_g$ is the local gas number density and $L$ is the loss function depending solely on the kinetic energy. Then, replacing
$\mu$ with the new variable $(1-\mu^2)/B(s)$ reduces Equation~(\ref{transportEquation0}) to
\begin{equation}
\frac{\hat\mu}{n_g}\frac{\partial j}{\partial s}-\frac{\partial}{\partial E}(Lj) = 0,
\label{transportEquation}
\end{equation}
where $\hat\mu(\mu_i,s)$ denotes a function of position for given $\mu_i$, as determined by Equation~(\ref{cos_alpha}).

We point out that Equation~(\ref{transportEquation}) can be used as long as losses do not result in a substantial scattering
of a CR particle along its path \citep[otherwise, a term describing the pitch-angle diffusion should be added,
see][]{Morfill1976}. Indeed, such losses can be treated as a sequence of instantaneous small reductions of the energy,
occurring during individual collision events with gas particles (and leading to the corresponding decrease of the Larmor
radius). This treatment is justified since both the time and the length scales of the collision interactions are
incomparably smaller than the respective gyration scales of a CR particle. The assumption of a negligible scattering is well
applicable to the ionization interactions, dominating the energy losses by non-relativistic protons \citep[see,
e.g.,][]{Ginzburg1964book,Padovani2009}. In the range of $10^5$~eV$\lesssim E\lesssim5\times10^8$~eV, the ionization loss
function is accurately approximated by
\begin{equation}
L(E) = L_*\left(\frac{E}{E_*}\right)^{-d},
\label{lossFunction}
\end{equation}
with $d \approx 0.82$, $L_* \approx 1.4\times10^{-14}$~eV~cm$^2$ and $E_*=10^5$~eV \citep[][]{Padovani2018}.

By multiplying Equation~\eqref{transportEquation} with $L(E)$, we obtain a general solution $jL=\Psi(x+y)$. Here, $\Psi$ is
an arbitrary function with
\begin{equation}
x(\mu,s) = \int^s \frac{n_g(s')\:ds'}{\hat\mu(\mu_i,s')}\;; \quad y(E) = \int^E \frac{dE'}{L(E')}\;,
\label{xAndy}
\end{equation}
where $\mu_i(\mu, s)=\pm\sqrt{1-(1-\mu^2)/\tilde B(s)}$ is the inverse of Equation~(\ref{cos_alpha}). This means that $jL$
is conserved along lines of constant $x+y$. In this section we are assuming a single-peaked magnetic field profile, as in
Figure~\ref{introFig}(b), and all quantities refer to forward-moving particles propagating toward the peak. Hence, the
relation to the isotropic interstellar spectrum $j_i=vn_i$ is given by
\begin{equation}
j_+(\mu,E,s)L(E) = \frac12j_i(E_i)L(E_i).
\label{conservedQuantity}
\end{equation}
The relation between energy $E_i$ at the cloud boundary and energy $E$ at position $s$ is obtained from
Equation~\eqref{xAndy},
\begin{equation}
E_i^{1+d} = E^{1+d} + (1+d)L_*E_*^d N_+\;.
\label{Ei}
\end{equation}
The latter is determined by
\begin{equation}
N_+(\mu,s) = \int_0^{s} \frac{n_g(s')\:ds'}{\sqrt{1-\tilde b(s',s)(1-\mu^2)}}\;,
\label{Iequation}
\end{equation}
which is the actual column density traversed by a forward-moving particle on its helical trajectory, depending on $\tilde
b(s', s) = B(s')/B(s) \leq 1$.

The ionization rate at position $s$ due to forward-moving particles is
\begin{equation}
\zeta_+(s) = \int_0^1d\mu\int_0^\infty j_+(\mu,E,s) \sigma_{\rm ion}(E)\:dE,
\label{zetaz}
\end{equation}
where $\sigma_{\rm ion}(E)$ is the ionization cross section. We notice that the mean energy $\varepsilon$ lost by a CR
particle per ionization event is practically independent of $E$ \citep[][]{Padovani2009}.  This yields a simple relation,
\begin{equation}\label{L_vs_sigma}
L(E)\approx \varepsilon\sigma_{\rm ion}(E).
\end{equation}
Therefore, we can use Equations \eqref{conservedQuantity}, \eqref{zetaz}, and \eqref{L_vs_sigma} to write
\begin{equation}
\zeta_+(s) =  \frac1{2\varepsilon}\int_0^1 d\mu \int_0^\infty j_i(E_i)L(E_i)\:dE,
\label{zetaU}
\end{equation}
with $E_i(E,\mu,s)$ given by Equation \eqref{Ei}.

To continue further, we generally need to assume an explicit form for $n_{\rm g}(s)$ and $B(s)$. However, we can also
consider two limiting cases, which are determined by the behavior of $B(s)$ and provide exact lower and upper bounds on
$\zeta_+$. The {\it lower bound} (L) occurs when $\tilde b(s', s) = 1$ for $0 < s' < s$. In this case Equations \eqref{Ei},
\eqref{Iequation}, and \eqref{zetaU} show that $\zeta_+$ is the same as if $B = B_i$ throughout the whole cloud. The {\it
upper bound} (U) occurs if $\tilde b(s', s) = 0$ for $0 < s' < s$. In this case, the ionization rate due to forward-moving
particles is increased relative to the constant-field case, because the CRs have accrued less column between $s' = 0$ and
$s' = s$. Thus, the ionization rate by forward-moving particles is always limited in the range of
\begin{equation*}
\zeta_{\rm L}\leq\zeta_+\leq\zeta_{\rm U}\;.
\end{equation*}
Physically, the lower bound $\zeta_{\rm L}$ represents propagation of CRs along a constant magnetic field, where both the
mirroring and focusing are absent, and therefore this is our {\it reference value} of the ionization rate. The upper bound
$\zeta_{\rm U}$ reflects an extreme situation of all CRs having zero pitch angles. Hence, the relative increase of the
ionization rate is conveniently quantified by the ratio $\mathcal{R}\equiv \zeta_+/\zeta_{\rm L}$, bounded between unity and
$\zeta_{\rm U}/\zeta_{\rm L}$.

In order to calculate the value of $\mathcal{R}_{\rm max}=\zeta_{\rm U}/\zeta_{\rm L}$, let us consider a typical model
spectrum of the interstellar CRs, approximately described by a power-law dependence for the non-relativistic energy range
\citep[][]{Ivlev2015b},
\begin{equation}
j_i(E) = j_*\left(\frac{E}{E_*}\right)^{-a}.
\label{JIS}
\end{equation}
We expect the results to be valid at column densities where Equation~(\ref{lossFunction}) is applicable, i.e., where all
particles with $E_i \lesssim 10^5$~eV have been attenuated, but particles with $E_i\gtrsim 5\times 10^8$~eV are not
attenuated significantly. This corresponds to column densities of roughly $10^{19}$~cm$^{-2}$ to $10^{25}$~cm$^{-2}$
\citep[][]{Padovani2018}.

If the value of the spectral index $a$ is sufficiently small, $a <1-d\approx0.2$, then the integral over $E$ in
Equation~\eqref{zetaU} is dominated by larger $E$. In practice, this means that $\zeta_+(s)$ remains approximately constant
as long as the column density is smaller than the stopping range for $E\approx 5\times10^8$~eV, where a crossover to the
relativistic spectrum $j_i\propto E^{-2.7}$ occurs \citep[][]{Ivlev2015b}. This stopping range nicely coincides with the
upper limit of column densities where Equation~(\ref{lossFunction}) is still applicable. Therefore, attenuation of CRs with
$a < 1-d$ does not (substantially) affect the value of $\zeta_+$ at column densities $\lesssim10^{25}$~cm$^{-2}$, and
effects of the mirroring and focusing for such interstellar spectra cancel out, as discussed in Section~\ref{noLoss}.

Thus, below we consider interstellar spectra with $a > 1-d\approx0.2$, for which the ionization at $s$ is dominated by lower
CR energies (viz., by the energies for which the stopping range is of the order of the column density at $s$). By
substituting Equations~(\ref{lossFunction}), (\ref{Ei}), and (\ref{JIS}) in Equation~(\ref{zetaU}), we obtain for the lower
ionization bound (reference value)
\begin{equation}\label{zetaI}
\zeta_{\rm L}(s) = \int_0^1K(\mu,s)d\mu,
\end{equation}
where
\begin{eqnarray}
K(\mu,s) = \frac{j_*L_*}{2\varepsilon}\nonumber\hspace{3.5cm}\\
\times\int_0^\infty\left[\left(\frac{E}{E_*}\right)^{1+d}+(1+d)\frac{L_*}{E_*}\frac{N_{\rm eff}}{\mu}\right]^{-\frac{a+d}{1+d}}dE,
\label{Kequation}
\end{eqnarray}
while for the upper bound we have
\begin{equation}\label{zetaII}
\zeta_{\rm U}(s) = K(1,s).
\end{equation}
Here, $N_{\rm eff}(s) = \int_0^{s}n_g(s') ds'$ is the so-called {\it effective column density}, measured along a field line
\citep[see, e.g.,][]{Padovani2018}. By making the substitution $E' = \mu^{\frac{1}{1+d}}E$ into Equation~(\ref{Kequation}),
one can easily show that
\begin{equation}
K(\mu,s) = \mu^\frac{a+d-1}{1+d} K(1,s).
\end{equation}
Thus, performing the integration over $\mu$ in Equation~(\ref{zetaI}), we find that $\zeta_{\rm U}/\zeta_{\rm L}$ does not
depend on $s$ and is equal to
\begin{equation}
\mathcal{R}_{\rm max}=1+\frac{a + d-1}{1+d}\;.
\label{fancyR}
\end{equation}
We conclude that, for any density profile and a single-peaked magnetic field, a combined effect of the mirroring and
focusing is able to increase the ionization rate (relative to the reference value) by a factor not larger than
$\mathcal{R}_{\rm max}$. For realistic values of the spectral index $a\lesssim1$, we have $\mathcal{R}_{\rm
max}\lesssim1.5$; this factor naturally tends to unity when $a=1-d$.

Equation~(\ref{fancyR}) becomes increasingly inaccurate above a column density of $\sim10^{25}$~cm$^{-2}$, corresponding to the stopping range of CR with energies where Equation~(\ref{lossFunction}) is no longer applicable.

\subsection{Contribution of the Mirrored Particles}
\label{mirroringSection}

In the previous section, we only considered the ionization by incoming particles. Now we will estimate the contribution from
the mirrored, backward-moving particles.  We expect this to be less than the ionization rate by the forward-moving
particles, because the mirrored particles have traveled through a larger column.

Let $\zeta_-(s)/\zeta_+(s)$ be a ratio of the ionization due to the mirrored particles to the ionization due to the
forward-moving particles. Calculation of this ratio is difficult in general. To make the problem tractable and estimate the
magnitude of the effect, we consider a model for the increase of magnetic field and gas density within the cloud, in which
we let
\begin{equation}
\tilde B(s) = \left(\frac{s}{H}\right)^p; \quad n_g(s) = n_{g*} \left(\frac{s}{H}\right)^q,
\label{Bandn}
\end{equation}
with $H$ being a relevant spatial scale and $p$ and $q$ positive. Even though Equation~(\ref{Bandn}) yields an unphysical
behavior of $B(s)$ for $s \lesssim H$, this region contributes negligibly compared to the region of interest ($s \gg H$),
where the mirroring and focusing effects are strong. In Appendix~\ref{mirrorFluxCalcs} we calculate $\zeta_+(s)$ and
$\zeta_-(s)$ and demonstrate that their ratio does not depend on $s$, and is characterized by the spectral index $a$ and the
ratio
\begin{equation}\label{r}
r = \frac{q+1}{p}-1.
\end{equation}
\begin{figure}[htp]
\centering
\includegraphics[width=.5\textwidth]{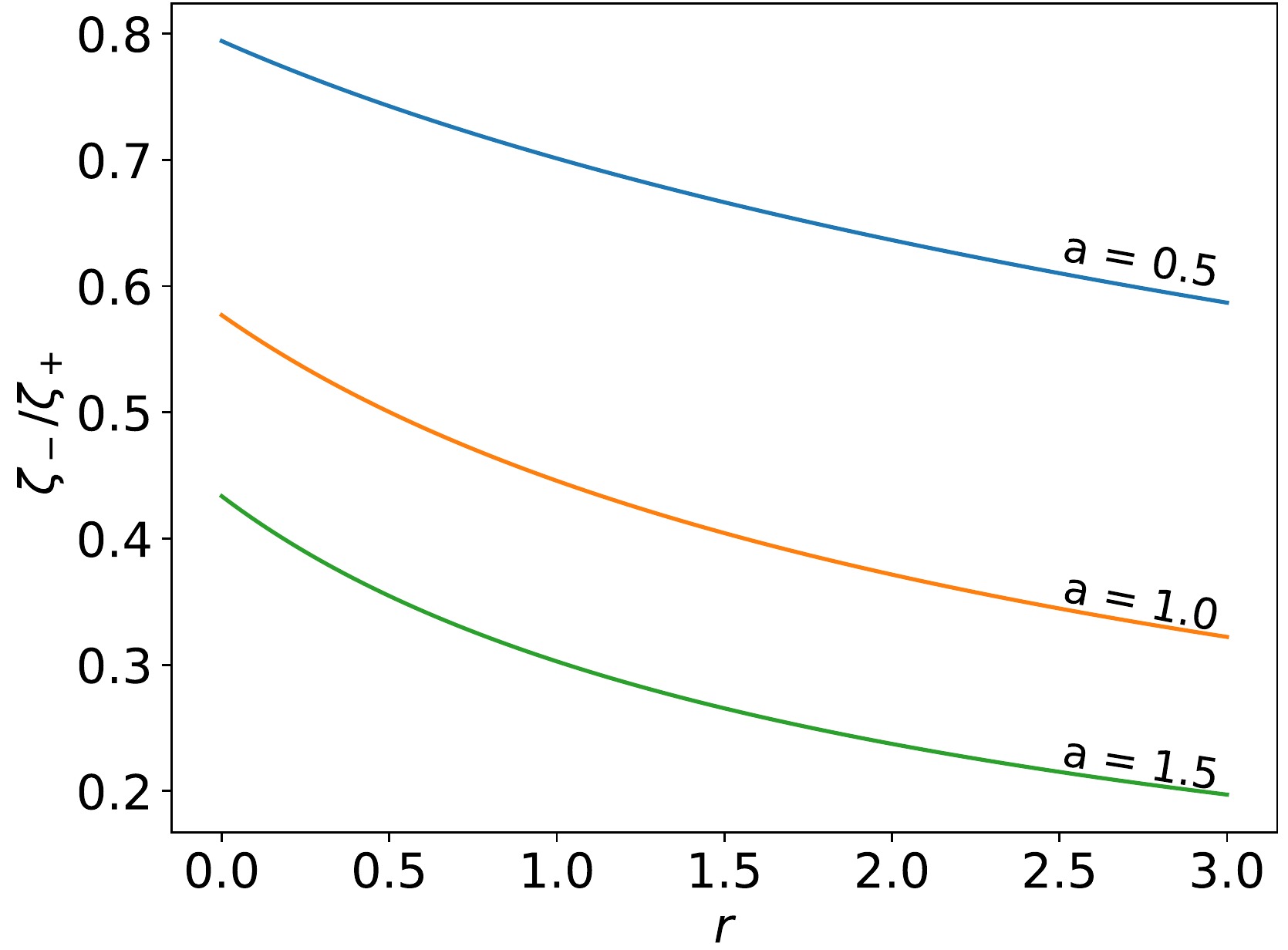}
\caption{Ratio of ionization rates due to the mirrored (backward-moving) and forward-moving CRs, $\zeta_-/\zeta_+$, plotted
versus the parameter $r$, Equation~(\ref{r}). The curves represent different values of the spectral index $a$ of interstellar
CRs, Equation~(\ref{JIS}).}
\label{mirroredFlux}\vspace{.1cm}
\end{figure}
Figure \ref{mirroredFlux} shows $\zeta_-/\zeta_+$ as a function of $r$, plotted for different values of $a$. The behavior is
easy to understand: it is a decreasing function, since larger $r$ implies smaller $p$ or larger $q$ and hence more column
between $s$ and the mirror point (relative to the column between $0$ and $s$). The constraint that $p$ and $q$ are positive
means that $r> -1$; we have $r\to-1$ for a sharply increasing field (i.e., for a vanishing column between $s$ and the mirror
point), which implies $\zeta_-\to\zeta_+$. Larger values of $a$ cause this decrease to happen more rapidly, as a harder
spectrum means that the ionization rate falls off faster with increased column depth.

Under the assumption of magnetic flux freezing during the cloud collapse, we would expect $\frac12 \leq p/q \leq \frac23$
\citep[][]{Crutcher2012}. Then we can write $r = r'+1/p$, where $\frac12 \leq r' \leq 1$.

\subsection{Ionization in magnetic pockets}
\label{pockets}

As discussed in Section~\ref{multipleMaxima}, the CR density can be drastically reduced at local field minima -- magnetic
pockets, where particles experience {\it defocusing} relative to the lower peak $B_l$, but the mirroring has no effect on
their motion. This fact has profound implications for the ionization.

Let us first assume that the attenuation in a pocket is negligible compared to that in the cloud -- we call such pockets
``localized''. The ionization rate by forward-moving particles, $\zeta_+^{\rm pock}$, can be calculated from
Equation~\eqref{zetaU} where $\sqrt{1 - B/B_l}$ is used for the lower limit of the integral over $\mu$ (see
Section~\ref{multipleMaxima}). The upper ionization bound, $\zeta_{\rm U}^{\rm pock}(s)$, for which we set $\tilde b=0$ in
Equation~(\ref{Iequation}), is readily obtained by following the same steps as in Section~\ref{FMP}. This yields
\begin{equation}\label{zeta_pockII}
{\rm Localized:}\quad\frac{\zeta_{\rm U}^{\rm pock}}{\zeta_{\rm L}}= \mathcal{R}_{\rm max}\left(1 -
\sqrt{1-\frac{B}{B_l}}\;\right),
\end{equation}
where the normalization is by the reference value $\zeta_{\rm L}$ and $\mathcal{R}_{\rm max}\equiv\zeta_{\rm U}/\zeta_{\rm
L}$ is given by Equation~(\ref{fancyR}). To derive the lower bound, representing the case of a constant field $B_l$ outside
the pocket, we notice that now $\tilde b=B_l/B\geq1$. Substituting Equation~(\ref{Iequation}) in Equation~\eqref{zetaU} and
introducing a new variable $x=(1-\mu^2)B_l/B$, after simple transformations we obtain
\begin{equation}\label{zeta_pockI}
\frac{\zeta_{\rm L}^{\rm pock}}{\zeta_{\rm L}}= \frac12\mathcal{R}_{\rm max}\frac{B}{B_l}\int_0^1
\frac{(1-x)^{\frac12(\mathcal{R}_{\rm max}-1)}}{\sqrt{1-(B/B_l)x}}\:dx.
\end{equation}
As expected, this yields $\zeta_{\rm L}^{\rm pock}\to\zeta_{\rm L}$ when $B\to B_l$;
for deep regions in the pockets, where $B/B_l$ is small, from Equations~(\ref{zeta_pockII}) and (\ref{zeta_pockI}) we derive
$\zeta_{\rm U}^{\rm pock}/\zeta_{\rm L}^{\rm pock}\approx\frac12(\mathcal{R}_{\rm max}+1)$. The latter shows that the
relative range of possible ionization rates in a (deep) pocket is half of that in the single-peaked case
(Section~\ref{FMP}).

Concerning the ionization by the mirrored particles, we notice that the value of $\zeta_-/\zeta_+$ is unaffected by the
presence of a localized pocket, since the latter has largely the same effect on the pitch angle of both mirrored and
forward-moving particles. Hence, their contribution can be evaluated using the results of Section~\ref{mirroringSection}.

We can also consider the opposite situation, where the column density of a magnetic pocket is much larger than that between
the edge of the cloud and the pocket. When calculating the ionization rate by forward-moving particles in such ``global''
pockets, the column accrued exterior to the peak $B_l$ can be ignored, i.e., one can assume that interstellar CRs directly
enter the pocket. In this case, the lower ionization bound would be where the magnetic field in the pocket remains at the
peak value $B_l$ up to position $s$, where it drops to the value $B(s)$. This is physically equivalent to the lower bound
derived for a localized pocket, and the ionization rate is therefore given by Equation~(\ref{zeta_pockI}). For the upper
bound, we assume that the magnetic field in the pocket decreases from $B_l$ to $B$ right at $s_l$. The corresponding
ionization rate can be derived from Equation~(\ref{zetaI}) (which is the lower bound for the single-peaked field) where,
again, the lower limit of integration is replaced with $\sqrt{1-B/B_l}$. We obtain that the ionization is reduced by a
factor of
\begin{equation}
{\rm Global:}\quad \frac{\zeta_{\rm U}^{\rm pock}}{\zeta_{\rm L}}=1-\left(1-\frac{B}{B_l}\right)^{\frac12
\mathcal{R}_{\rm max}}.
\label{basin}
\end{equation}
From Equations~(\ref{zeta_pockI}) and (\ref{basin}) it follows that the upper and lower ionization bounds in global pockets
coincide when $B\to B_l$, and tend to the reference value $\zeta_{\rm L}$. The latter is easy to understand, since in this
case CRs propagate along a quasi-constant magnetic field. In deep regions the difference between local and global pockets
disappear and the upper-bound reduction factors, given by Equations~(\ref{zeta_pockII}) and (\ref{basin}), tend to the same
value of $\zeta_{\rm U}^{\rm pock}/\zeta_{\rm L}\approx\frac12\mathcal{R}_{\rm max}(B/B_l)\ll1$. From
Equation~(\ref{zeta_pockI}) we infer that the lower-bound reduction in this limit is a factor of $\frac12(\mathcal{R}_{\rm
max}+1)$ smaller.

The ionization by mirrored particles is reduced significantly in a global pocket. Indeed, forward-moving CRs entering the
pocket from one side propagate without mirroring until (at least) the other side. A mirrored particle at a given position
accrues on average the column density of the entire global pocket before returning back. Thus, the contribution of mirrored
particles can be reasonably neglected.

\section{Discussion and conclusions}
\label{discussion}

In order to draw general conclusions about the net effect of the magnetic mirroring and focusing, let us start with the
single-peaked field profile, sketched in Figure~\ref{introFig}(b), and consider two characteristic situations. One is when
the ionization occurs sufficiently away from the center of a cloud, where $B(s)$ is still substantially smaller than the
peak value $B_p$. CRs originating from the other side of the cloud are negligible in this case, both due to the strong
attenuation and the narrow range of initial pitch angles that allow the particles to overcome the mirroring. Based on the
results of Sections~\ref{FMP} and \ref{mirroringSection}, we immediately obtain that the relative increase of the ionization
rate with respect to the reference value $\zeta_{\rm L}$ (representing a constant field strength) is equal to
$\mathcal{R}(1+\zeta_-/\zeta_+)$, where the upper bound for $\mathcal{R}$ is given by Equation~(\ref{fancyR}) and the lower
bound is unity. The other situation occurs near the cloud center, where the contribution of the mirrored particles is no
longer important, and the ionization is due to CRs coming from both sides of the cloud. Then the relative increase of the
ionization rate is simply $\mathcal{R}$.  We have shown that for realistic values of the spectral index $a$, the total
relative increase does not exceed a factor of 1.5--2.

In the presence of multiple magnetic peaks, illustrated in Figure~\ref{introFig}(c), these conclusions remain unchanged
everywhere except for the regions of local field minima -- magnetic pockets, where the ionization can be decreased
drastically. In Section~\ref{pockets} we show that the ionization decrease is described by a reduction factor for which we
also consider two characteristic situations: localized pockets with column small compared to that between the pocket and the
cloud edge, and the opposite situation of ``global'' pockets. In the former case, for practical purposes one can employ
Equation~(\ref{zeta_pockII}) for the reduction factor; for global pockets, the reduction is given by Equation \eqref{basin}.
For pockets with intermediate column densities it is not crucial which formula is used, since they differ at most by a
factor of $\leq R_{\rm max}$.

Calculations of the ionization rate in dense molecular clouds and circumstellar disks
inevitably contain significant intrinsic uncertainties. These are primarily associated with poor knowledge of the gas
distribution and the configuration of magnetic field lines in the cores, both leading in uncertainties in the effective
column density $N_{\rm eff}$ and, hence, in the reference ionization rate $\zeta_{\rm L}(N_{\rm eff})$. The analysis
presented above shows that, compared to these uncertainties, the variations due to the CR mirroring and focusing are
negligible and therefore can be safely neglected -- as long as the field has a single-peaked profile. Within the magnetic
pockets, the derived analytical formulas should be used to accurately calculate the relative decrease of the ionization rate
in these regions.

The results of this paper were derived for effective column densities of up to $N_{\rm eff}\sim10^{25}$~cm$^{-2}$.
Remarkably, \citet[][]{Padovani2018} have recently shown that at $N_{\rm eff}\gtrsim3\times10^{25}$~cm$^{-2}$ the CR
ionization is driven by indirect processes, mediated by secondary photons, and then the magnetic field plays no role at all.
This allows us to conclude that the mirroring and focusing do not significantly affect the ionization outside the magnetic
pockets at any column density.

We note that \citet[][]{Padovani2011} performed a numerical study of the CR mirroring and focusing for a specified
density and magnetic field distribution, valid for a molecular cloud core in magnetostatic equilibrium. They also reached
the conclusion that there was a near-cancelation of the magnetic mirroring and focusing terms. In detail, however, there
appears to be a minor discrepancy with the present results -- specifically, they find some reduction in the ionization rate
due to the combined effect of mirroring and focusing, whereas we find a slight increase in the ionization rate. This
discrepancy arises from their overestimate of the reduction of CR flux by mirroring (by the factor of $1-\sqrt{1-1/\tilde
B}$ rather than $1/{\tilde B}$), and by their assumption that CRs lose all their kinetic energy in approaching the mirror
point. The latter assumption was relaxed by \citealt[][]{Padovani2013}, who studied the propagation of CRs along the
magnetic field for a rotating collapsing core with different initial conditions (mass-to-flux ratio, angle between the mean
magnetic field direction and the rotation axis). The severe reduction of the CR ionization rate, reported in this paper for
certain regions, was attributed to the stronger effect of mirroring with respect to focusing.  In fact, preliminary
analysis of their simulation data shows this strong reduction to be due to the presence of magnetic pockets, which is fully
consistent with the results presented in Section~\ref{pockets}.

To summarize, the effects of magnetic mirroring and focusing on the local CR density practically cancel each other out if
there are no magnetic pockets. This implies we can safely use available numerical results for CR ionization (calculated
neglecting these magnetic effects) and simply assume the CR propagation along field lines. For situations where magnetic
pockets are present, the ionization rate can be greatly reduced. We obtained simple expressions allowing accurate
calculation of the ionization in localized or global pockets. In a separate paper, we plan to investigate conditions under
which magnetic pockets may form, and study their effect on non-ideal MHD processes occurring in dense cores and disks.
Also, we will analyze the role of the CR diffusion along the magnetic field, caused by CR collisions with particles of the
medium or by their (resonant and non-resonant) interaction with fluctuating field.

\section*{Acknowledgements}

MP acknowledges funding from the European Unions Horizon 2020 research and innovation programme under the Marie
Sk\l{}odowska-Curie grant agreement No~664931.

\appendix

\section{Appendix A\\ Ionization due to the mirrored particles}
\label{mirrorFluxCalcs}

Let $N_+(\mu, s)$ and $N_-(\mu, s)$ be the column densities accrued at position $s$ by, respectively, the forward-moving and
mirrored particles having the local pitch-angle cosine of $\pm\mu$. The corresponding ionization rates $\zeta_+(s)$ and
$\zeta_-(s)$ can be derived from Equation~\eqref{zetaU}.

Substituting Equation~(\ref{Ei}) in Equation~\eqref{zetaU} and introducing new variable $E' = \left[(1+d)(L_*/E_*)N_\pm
\right]^{-\frac1{1+d}} E$, we decouple $N_\pm$ from the resulting integrals over energy. For our purpose it is convenient to
replace the integration over $\mu$ with the integration over $\mu_i$, by using Equation~(\ref{dmu}). For locations in the
cloud where $B(s) \gg B_i$ one can set $\mu_i\approx 1$ and $1-\mu_i^2 \approx \alpha_i^2$, so that $\mu\approx
\pm\sqrt{1-\tilde B(s)\alpha_i^2}$. Using the initial energy spectrum~(\ref{JIS}), we obtain
\begin{equation}\label{TInFront}
\zeta_\pm(s)=T(s)\int_0^{1/\tilde B(s)}N_\pm^{1-\mathcal{R}_{\rm max}}\frac{d\alpha_i^2}{\sqrt{1-\tilde B(s)\alpha_i^2}}\;,
\end{equation}
where $T(s)$ is a common pre-factor and $\mathcal{R}_{\rm max}$ is given by Equation~(\ref{fancyR}). In the same
approximation, from Equation~(\ref{Iequation}) we obtain
\begin{equation*}
N_+(\alpha_i,s) = \int_0^{s} \frac{n_g(s')\: ds'}{\sqrt{1-\tilde B(s')\alpha_i^2}}\;.
\end{equation*}
Then, we assume $\tilde B(s)$ and $n_g(s)$ to be given by Equation \eqref{Bandn}, and introduce $x = (s/H)^p\alpha_i^2$ and
$x' = (s'/H)^p\alpha_i^2$. This yields
\begin{equation}
N_+ = \frac{n_{g*}H}{p\alpha_i^{2(r+1)}}\int_0^{x} \frac{x'^r\:dx'}{\sqrt{1-x'}}\;,
\label{N1}
\end{equation}
where $r = (q+1)/p-1$. Similarly, we write
\begin{equation}
N_- = N_+ + \frac{2n_{g*}H}{p\alpha_i^{2(r+1)}}\int_{x}^1 \frac{x'^r\:dx'}{\sqrt{1-x'}}\;,
\label{N2}
\end{equation}
where we used the fact that $x' = 1$ is the mirror point. Finally, substituting Equations~\eqref{N1} and \eqref{N2} in
Equation~\eqref{TInFront}, we derive
\begin{equation*}
\zeta_+(s) = T'(s) \int_0^1\left[\frac{1}{x^{r+1}}\int_{0}^x \frac{x'^r\:dx'}{\sqrt{1-x'}}
\right]^{1-\mathcal{R}_{\rm max}}
\frac{dx}{\sqrt{1-x}}\;,
\end{equation*}
and
\begin{equation*}
\zeta_-(s) = T'(s) \int_0^1 \left[\frac{1}{x^{r+1}}\left(\int_{0}^x \frac{x'^r\:dx'}{\sqrt{1-x'}}
+ 2\int_x^1 \frac{x'^r\:dx'}{\sqrt{1-x'}}\right)\right]^{1-\mathcal{R}_{\rm max}}\frac{dx}{\sqrt{1-x}}\;,
\end{equation*}
where $T'(s)$ is a (new) common pre-factor. We conclude that the ratio $\zeta_-/\zeta_+$ does not depend on $s$. The ratio
is obviously smaller than unity (since $\mathcal{R}_{\rm max}(a)>1$), and is a function of parameters $r$ and $a$.

\bibliographystyle{apj}
\bibliography{mirroringApJ4}

\begin{thebibliography}{}
\expandafter\ifx\csname natexlab\endcsname\relax\def\natexlab#1{#1}\fi

\bibitem[{{Allen} {et~al.}(2003){Allen}, {Li}, \& {Shu}}]{Allen2003}
{Allen}, A., {Li}, Z.-Y., \& {Shu}, F.~H. 2003, \apj, 599, 363

\bibitem[{{Cesarsky} \& {V\"olk}(1978)}]{Cesarsky1978}
{Cesarsky}, C.~J., \& {V\"olk}, H.~J. 1978, \aap, 70, 367

\bibitem[{{Chandran}(2000)}]{Chandran2000}
{Chandran}, B.~D.~G. 2000, \apj, 529, 513

\bibitem[{{Chen}(1974)}]{Chen1984book}
{Chen}, F.~F. 1974, {Introduction to plasma physics} (New York: Plenum Press)

\bibitem[{{Crutcher}(2012)}]{Crutcher2012}
{Crutcher}, R.~M. 2012, \araa, 50, 29

\bibitem[{{Desch} {et~al.}(2004){Desch}, {Connolly}, \&
  {Srinivasan}}]{Desch2004}
{Desch}, S.~J., {Connolly}, Jr., H.~C., \& {Srinivasan}, G. 2004, \apj, 602,
  528

\bibitem[{{Ginzburg} \& {Syrovatskii}(1964)}]{Ginzburg1964book}
{Ginzburg}, V.~L., \& {Syrovatskii}, S.~I. 1964, {The Origin of Cosmic Rays}
  (Oxford: Pergamon)

\bibitem[{{Ivlev} {et~al.}(2018){Ivlev}, {Dogiel}, {Chernyshov}, {Caselli},
  {Ko}, \& {Cheng}}]{Ivlev2018}
{Ivlev}, A.~V., {Dogiel}, V.~A., {Chernyshov}, D.~O., {et~al.} 2018, \apj, 855,
  23

\bibitem[{{Ivlev} {et~al.}(2015){Ivlev}, {Padovani}, {Galli}, \&
  {Caselli}}]{Ivlev2015b}
{Ivlev}, A.~V., {Padovani}, M., {Galli}, D., \& {Caselli}, P. 2015, \apj, 812,
  135

\bibitem[{{Joos} {et~al.}(2012){Joos}, {Hennebelle}, \& {Ciardi}}]{Joos2012}
{Joos}, M., {Hennebelle}, P., \& {Ciardi}, A. 2012, \aap, 543, A128

\bibitem[{{Ko}(1992)}]{Ko1992}
{Ko}, C.-M. 1992, \aap, 259, 377

\bibitem[{{Lazarian}(2014)}]{Lazarian2014}
{Lazarian}, A. 2014, \ssr, 181, 1

\bibitem[{{Li} {et~al.}(2013){Li}, {Krasnopolsky}, \& {Shang}}]{Li2013}
{Li}, Z.-Y., {Krasnopolsky}, R., \& {Shang}, H. 2013, \apj, 774, 82

\bibitem[{{Mestel} \& {Spitzer}(1956)}]{Mestel1956}
{Mestel}, L., \& {Spitzer}, Jr., L. 1956, \mnras, 116, 503

\bibitem[{{Morfill} {et~al.}(1976){Morfill}, {V\"olk}, \& {Lee}}]{Morfill1976}
{Morfill}, G.~E., {V\"olk}, H.~J., \& {Lee}, M.~A. 1976, \jgr, 81, 5841

\bibitem[{{Padoan} \& {Scalo}(2005)}]{Padoan2005}
{Padoan}, P., \& {Scalo}, J. 2005, \apjl, 624, L97

\bibitem[{{Padovani} \& {Galli}(2011)}]{Padovani2011}
{Padovani}, M., \& {Galli}, D. 2011, \aap, 530, A109

\bibitem[{{Padovani} {et~al.}(2009){Padovani}, {Galli}, \&
  {Glassgold}}]{Padovani2009}
{Padovani}, M., {Galli}, D., \& {Glassgold}, A.~E. 2009, \aap, 501, 619

\bibitem[{{Padovani} {et~al.}(2013){Padovani}, {Hennebelle}, \&
  {Galli}}]{Padovani2013}
{Padovani}, M., {Hennebelle}, P., \& {Galli}, D. 2013, \aap, 560, A114

\bibitem[{{Padovani} {et~al.}(2018){Padovani}, {Ivlev}, {Galli}, \&
  {Caselli}}]{Padovani2018}
{Padovani}, M., {Ivlev}, A.~V., {Galli}, D., \& {Caselli}, P. 2018, \aap, 614,
  A111

\bibitem[{{Parker}(1963)}]{Parker1963book}
{Parker}, E.~N. 1963, {Interplanetary dynamical processes} (New York:
  Interscience Publishers)

\bibitem[{{Price} \& {Bate}(2008)}]{Price2008}
{Price}, D.~J., \& {Bate}, M.~R. 2008, \mnras, 385, 1820

\bibitem[{{Skilling} \& {Strong}(1976)}]{Skilling1976}
{Skilling}, J., \& {Strong}, A.~W. 1976, \aap, 53, 253

\bibitem[{{Yamamoto}(2017)}]{Yamamoto2017book}
{Yamamoto}, S. 2017, {Introduction to Astrochemistry: Chemical Evolution from
  Interstellar Clouds to Star and Planet Formation} (Tokyo: Springer)

\bibitem[{{Zhao} {et~al.}(2018){Zhao}, {Caselli}, {Li}, \&
  {Krasnopolsky}}]{Zhao2018}
{Zhao}, B., {Caselli}, P., {Li}, Z.-Y., \& {Krasnopolsky}, R. 2018, \mnras,
  473, 4868

\end{thebibliography}

\end{document}